\documentclass[10pt]{article} 
\usepackage{lmodern} 
\usepackage[utf8,utf8x]{inputenc}
\usepackage[a4paper]{geometry}
\usepackage{amsfonts,amsmath,amssymb,latexsym,amsthm} 
\usepackage{graphics,graphicx}
\usepackage[normalsize]{caption}
\usepackage{algorithm}
\usepackage{algorithmic}

\def \no{\noindent}

\def \b{\big}
\def \m{\medskip}
\def \s{\smallskip}
\def \cE{\mathcal{E}}
\def \cM{\mathcal{M}}
\def \cO{\mathcal{O}}
\def \cP{\mathcal{P}}
\def \cR{\mathcal{R}}
\def \cS{\mathcal{S}}
\def \cW{\mathcal{W}}

\newtheorem{lem}{Lemma}

\newcommand{\bc}{\begin{center}}
\newcommand{\ec}{\end{center}}

\begin{document}

\title{A distributed wheel sieve algorithm based\\
on {\sl Scheduling by Multiple Edge Reversal}
}
\author{Christian Lavault$^*$ \and Gabriel A. L. Paillard \and Felipe M. G. Fran\c{c}a$^\dag$}
\date{\small 
$^*$ Laboratoire d’Informatique de Paris Nord (LIPN UMR 7030 CNRS)\\
Institut Galil\'ee Universit\'e Paris XIII Villetaneuse F93430\\
{\em E-mail:} {{\tt Gabriel.Paillard,Christian.Lavault}@lipn.univ-paris13.fr}\\[.3\baselineskip]
$^\dag$ Programa de Engenharia de Sistemas e Computa\c{c}\~ao\\
Universidade Federal do Rio de Janeiro COPPE/UFRJ\\
Caixa Postal 68511 Cep: 21941-972, Rio de Janeiro RJ, Brazil\\
{\em E-mail:} {\tt felipe@cos.ufrj.br}
}
\maketitle

\begin{abstract}
This paper presents a new distributed approach for generating all prime numbers in a given interval of integers. From Eratosthenes, who elaborated the first prime sieve (more than 2000 years ago), to the current generation of parallel computers, which have permitted to reach larger bounds on the interval or to obtain previous results in a shorter time, prime numbers generation still represents an attractive domain of research and plays a central role in cryptography. We propose a fully distributed algorithm for finding all primes in the interval $[2\ldots, n]$, based on the \emph{wheel sieve} and the SMER (\emph{Scheduling by Multiple Edge Reversal}) multigraph dynamics. Given a multigraph $\cM$ of arbitrary topology, having $N$ nodes, a SMER-driven system is defined by the number of directed edges (arcs) between any two nodes of $\cM$, and by the global period length of all ``arc reversals'' in $\cM$. The new prime number generation method inherits the distributed and parallel nature of SMER and requires at most $n + \lfloor \sqrt{n}\rfloor$ time steps. The message complexity achieves at most $n\Delta_N + \lfloor \sqrt{n}\rfloor \Delta_N$, where $1\le \Delta_N\le N - 1$ is the maximal multidegree of $\cM$, and the maximal amount of memory space required per process is $\cO(n)$ bits.

\s \no {\bf Keywords:}\ Distributed Algorithms; Prime Numbers Generation; Wheel Sieve; 
Scheduling by Edge Reversal; Scheduling by Multiple Edge Reversal.
\end{abstract}

\section{Introduction}
This article takes up the generation of prime numbers smaller than a given bound $n$, by using the wheel sieve distributively. Wheel sieve algorithms can be very efficient to determine the primality
of integers which belong to a given finite interval $[2\ldots, n]$, for sufficiently large values of $n$ and when the test of primality is carried out on all numbers of the interval. The paper designs a fully distributed wheel sieve algorithm using scheduling by multiple edge reversal (SMER).

The main purpose of a parallelization of such kind of algorithm is to increase the bounds of the generation of prime numbers, and to reach these bounds in a shorter execution time. The first parallelization of a sieve algorithm was realized in 1987 \cite{BOK}, who parallelized the sieve of Eratosthenes. This work was motivated by testing a new parallel machine (the \emph{Flex/32}), because this kind of algorithm is ideal to test the performances of a new architecture (of a sequential or parallel machine) as a benchmark.

\m The sieve of Eratosthenes was the first prime sieving algorithm, and it consists in eliminating all non prime numbers in the interval $[2\ldots, n]$. First, the algorithm takes the first number of the interval and generates all its multiples (by adding its own value to himself), which are thus eliminated. The next (non eliminated) number is the one (the next prime number) which sieves the interval, and this process is pursued until all intervals has been sieved. Various parallelizations of this algorithm can be found, e.g. in \cite{SORE1,SORE2}.

However, the main drawback of the practical sieve of Eratosthenes is clearly the fact that it imposes to go through all
the entries of the multiples of each number during the sieving process. For instance, if the current entry  corresponds 
to $p$, then any entry at locations $2p$, $3p$, $4p$ is changed to zero, and so on, until the stop criteria is reached, i.e., $p^{2} > n$. The basic sieve of Eratosthenes proceeds in the same way on any other entry. It is easy to see that some numbers will be generated more than once, for example $6$ is generated twice (from $2$ and $3$), and $12$ is generated three times (from $2,3$ and $4$). The entries that are already zeros are left unchanged, but each entry must nevertheless be checked throughout the sieving process.

\m The main idea consists then in trying to prevent all numbers from being sieved ``too many times''. Sieving the
multiples of any given number more than once must be avoided, as much as possible. All efficient sieving algorithms are based on similar techniques. So, the complexity $\cO(n \ln\ln n)$ of the sieve of Eratosthenes may be somewhat improved by several clever arguments that are carried out by the above methods. Such sieve algorithms achieve a linear 
\cite{GRI,MAI,SORE1} or even a sublinear (step) complexity \cite{MAI,PRI1}.\par
So far, the best algorithm known is the ``wheel sieve'', designed in 1981 \cite{PRI1,PRI2}. It requires only 
$\cO(n/\log\log n)$ steps to find the set of primes in the interval $[2,\ldots ,n]$ (with $n > 4$), where each step is either for bookkeeping or an addition with integers at most $n$. Basically, the algorithm relies on the central result on the number of primes in arithmetic progressions. More precisely, Dirichlet's theorem states that if $a$, $b$ are coprime integers ($\gcd(a,b) := (a,b) = 1$) and $b > 0$, then the arithmetic progression $\b\{a, a+b, a+2b,\ldots\b\} %
= \b\{a \bmod b\b\}$ contains infinitely many primes~\cite[Thm.~15]{HAWR}. (See~\cite{CRAPO} for more details on the analysis of the wheel sieve algorithm.)

\m The paper presents a new kind of fully distributed algorithm that finds all primes by sieving in a given interval 
$[1\ldots, n]$, using the properties of the wheel sieve using the SMER \cite{PRI2}. Some other distributed algorithms generating all prime numbers can be found in \cite{COS1,COS2}, which use the properties of Dirichlet's theorem. 
In \cite{PAILLARD1} another kind of distributed prime number generation is presented, based only on scheduling by multiple edge reverse framework~\cite{VAL}.

In Sect.~2,  the wheel sieve algorithm is introduced. In Sect.~3 and 4, the framework of the \emph{scheduling by edge reversal} (SER) and the \emph{scheduling by multiple edge reversal} (SMER) mechanisms are both introduced. Sect.~5 is devoted to the design of our distributed  algorithm for sieving primes by using the SMER-based method applied to the wheel sieve. The worst-case complexity analysis of the algorithm is achieved in Sect.~6. 
The final Sect.~7 draws a short conclusion and offers some perspectives.   

\section{The Wheel Sieve}
The wheel sieve derived from Pritchard's algorithm \cite{PRI1} operates basically by generating a set of numbers that are not multiples of the first $k$ prime numbers. The sieve, applied on the resulting set from the wheel, eliminates the non prime numbers that remain in the set. This is the basic idea of the \emph{wheel} which were employed as a reduced residue class $\bmod(\Pi_{k})$, where $\Pi_{k}$ denotes the product of the first $k$ prime numbers \cite{PRI2}. $\cW_{k}$ denotes the $k$-th wheel, which is defined as 
\begin{equation}
\cR(x) = \b\{x\ /\ 1 \le y \le x\ \hbox{and}\ (y,x) = 1\b\},
\end{equation} 
where $x$ and $y$ are coprime numbers.

The sieve introduced by the wheel sieve consists basically, after having generated the next wheel $\cW_{k+1}$, in using the prime number $k + 1$ to sieve the new wheel, generating all its multiples and removing them from $\cW_{k+1}$. For more clarity this new set will denoted $\cS_{k+1}$. It is clear that after $\cS_{k+1}$ is obtained the algorithm proceeds to another sieving process, and eliminate the remaining composite numbers. The wheels are thus patterns that are repeated every $\Pi_{k}$ times. 

\m \begin{figure}[!ht]
\centering
\includegraphics[scale=0.35]{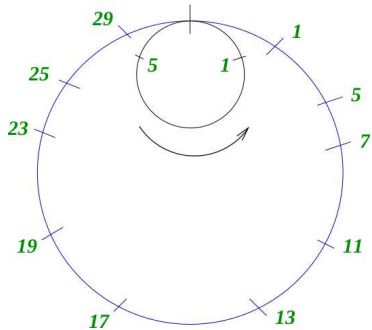} 
\caption{\em Example of the generation of a wheel $\cW_{k+1}$ starting from the preceding wheel $\cW_{k}$.}
\label{fig:Wheel1}
\end{figure}
In Fig.~\ref{fig:Wheel1} we use $\Pi_{2}$ in the first step of the wheel sieve as the product of the first two prime numbers $(2$ and $3)$ figured by the small circle; this generates all ``pseudo-primes'' numbers\footnote{Numbers that are not multiples of the first $k$ prime numbers.} between $1$ and the new bound contained in the new wheel $\cW_{3} = \cR(30)$, that is the actual bound $\Pi_{2}$ multiplied by the next prime $p_{3} = 5$. The next prime is the first number 
after $1$ that belongs to the interval being sieved \cite{PRI2} in the second wheel, which contains now the next value $5$.

\m \begin{figure}[!ht]
\centering
\includegraphics[scale=0.35]{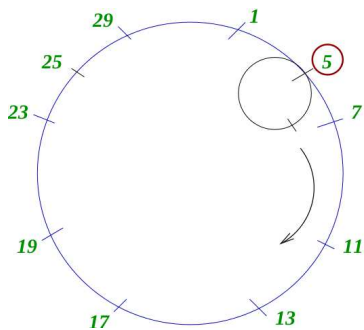}
\caption{{\em Generation of the new ``pseudo-prime'' numbers}}
\label{fig:Wheel2}
\end{figure}

\s The above Fig.~\ref{fig:Wheel2} shows that all the pseudo-prime numbers of the big wheel, that is 
$\{1,5,7,11,13,17,19,23,25,29\}$, are generated within the small wheel.

\newpage \begin{figure}[!ht]
\centering
\includegraphics[scale=0.35]{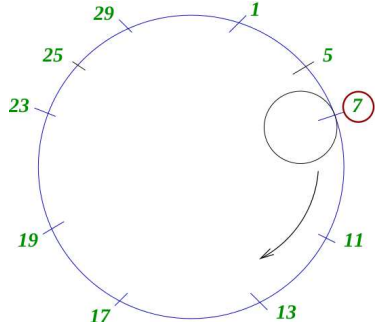}
\caption{\em Generation of another new pseudo-prime (number $7$).}
\label{fig:Wheel3}
\end{figure}

\s In Fig.~\ref{fig:Wheel3} the process of generating the big wheel is going on. Number $7$ is generated from the number $1$ of the small wheel, which can be interpreted as if we were ``rolling'' the small circle inside the big one. This means that starting from a wheel $\cW_{k}$, we can generate the next wheel $\cW_{k+1}$ in a graphical way. The points where the elements of the wheel $\cW_{k}$ touch the circle featuring $\cW_{k+1}$ are the new pseudo-primes. 
More precisely, $\cW_{k+1}$ is defined as
\begin{equation} \label{wheel}
\cW_{k+1} = \cW_{k} \cup \b\{x\Pi_{k} + y\ /\ x\in\{1,\ldots,p_{k+1}-1\}\ \hbox{and}\ y \in \cW_{k}\b\}.
\end{equation}

\m \begin{figure}[!ht]
\centering
\includegraphics[scale=0.35]{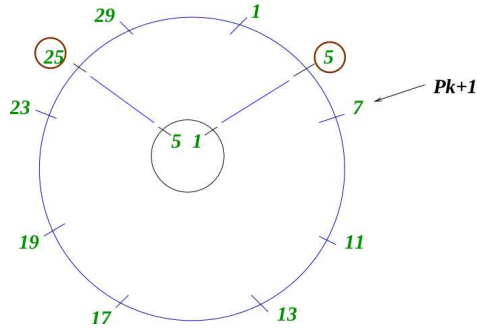}
\caption{\em The sieve being applied on the new wheel $\cW_{k+1}$ to generate $\cS_{k+1}$.}
\label{fig:Wheel4}
\end{figure}

\s Fig.~\ref{fig:Wheel4} shows the final phase of the wheel sieve, where the multiples of the previous $p_{k+1}$ (in that case, the number $5$) are eliminated from the set $\cR\b(\Pi_{3}\b)$. According to the definition of $\cW_{k+1}$ 
in Eq.~(\ref{wheel}), we also define
\begin{equation}
\cS_{k+1} = \cW_{k+1}\setminus \b\{y\times p_{k+1}\ /\ y\in \cW_{k+1}\b\}.
\end{equation}
The previous wheel $\cW_{k}$ is put in the center of the new wheel $\cS_{k+1}$ (See Fig.~\ref{fig:Wheel4}). Then drawing a radius from the center of the small circle containing each pseudo-prime number of this circle, each one of the prolongations of such radii touches the big circle at every pseudo-prime that will be eliminated in the new wheel $\cW_{k+1}$. Thus, the prime $p_{k+1}$ will be put in the set $\cP$ of all prime numbers.

\m In \cite{PAILLARD} a distributed version of the wheel sieve is proposed. It is implemented by using a message passing interface specification ({\emph{lam-mpi 7.0.6 library}})~\cite{MAI}. The time measurements of a sequential and
a distributed implementation of the wheel sieve are compared, together with a sequential and distributed implementation of the sieve of Eratosthenes. In \cite{PAILLARD2} a fully distributed version of the wheel sieve is also presented.

\section{Scheduling by Edge Reversal (SER)}
Consider a neighbourhood-constrained system composed by a set of \textit{processing elements} (PEs) and a set 
of \textit{atomic shared resources} represented by a connected directed graph $G = (V,E)$,  where $V$ is the set of PEs and $E$ the set of its directed edges (or arcs), stating the access topology (directed edges are henceforth refered to as arcs). 
The latter is defined in the following way: an arc exists between any two nodes \textit{if, and only if,} the two corresponding PEs share at least one atomic resource. SER works as follows: starting from any acyclic 
orientation $\omega$ on $G$, there is at least one \textit{sink} node, i.e., a node such that all its arcs are directed to itself; all sink nodes are allowed to operate while other nodes remain idle.

This obviously ensures mutual exclusion at any access made to shared resources by sink nodes. After operation, a sink node will reverse the orientation of its arcs, becoming a \textit{source} and thus releasing the access to resources to its neighbours. A new acyclic orientation is defined and the whole process is then repeated for the new set of sinks. 
Let $\tilde{\omega} = g(\omega)$ denote this greedy operation. SER can be regarded as the endless repetition of the application of $g(\omega)$ upon $G$. 

Assuming that $G$ is finite, it is easy to see that eventually a set of acyclic orientations will be repeated defining a period of length $P$. This simple dynamics ensures that no deadlocks or starvation will ever occur since in every acyclic orientation there exists at least one sink, i.e. one node allowed to operate. Also, it is proved that inside any period, every node operates exactly the same constant number of times (denoted $M$)~\cite{GAF2}.

\m SER is a fully distributed graph dynamics in which the sense of time is defined by its own operation, i.e., the synchronous behavior is equivalent to the case where every node in $G$ takes an identical amount of  time to operate and also an identical amount of time to reverse arcs. Another interesting observation to be made here is that any topology 
$G$ will have its own set of possible SER dynamics~\cite{VAL}. 

\m \begin{figure}[!ht]
\centering
\includegraphics[scale=0.35]{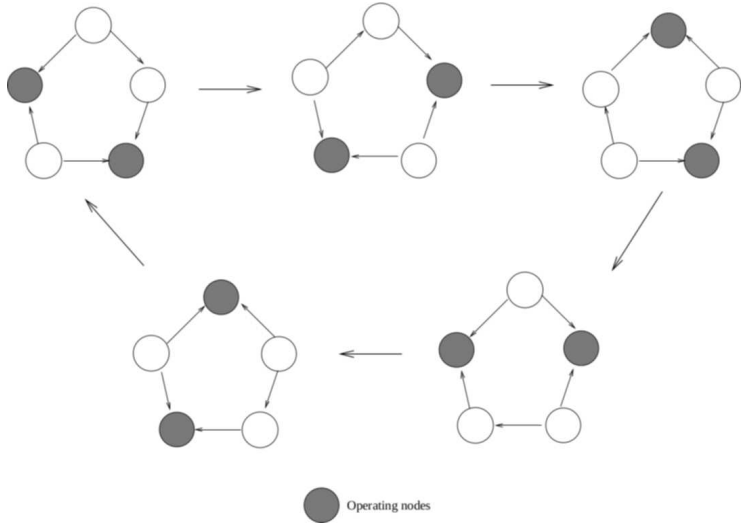}
\caption{\em SER dynamics for the Dining Philosophers under heavy load.}
\label{fig:dining}
\end{figure}

\s As an example of SER's applicability, consider Dijkstra's paradigmatic {\em Dining Philosophers Problem}~\cite{DJ} under heavy load, i.e., in the case philosophers are either ``hungry'' or ``eating'' (no ``thinking'' state). Such system can be represented by a set $\b\{P_{1},\ldots, P_{N}\b\}$ of $N$ PEs, in which each PE shares a resource both with its previous PE and its subsequent PE. Thus, taking the original configuration where $N =5 $ and setting an acyclic orientation over the $5$ nodes ring, the resulting SER dynamics where $P=5$ and $M=2$ is illustrated in Fig.~\ref{fig:dining}.

\section{Scheduling by Multiple Edge Reversal (SMER)}
SMER is a generalization of SER in which pre-specified access rates to atomic resources are imposed to processes in a distributed resource-sharing system represented by a multigraph $\cM = (V,\cE)$. In contrast with SER, multiple edges can exist between any two nodes $i$ and $j$ ($i,\, j\in V$) in the SMER dynamics: there can exist $e_{i,j}\ge 0$ undirected edges connecting nodes $i$ and $j$; such connected  nodes are called ``neighbours''. 

Let $r_i$ denote the ``reversibility'' of node $i$, as defined in \cite{FRA}. More precisely, reversibility $r_i$ is the number  of arcs that shall be reversed by $i$ towards each of its neighbouring nodes at the end of each operation step (access to  shared resources). Node $i$ is called a $r$-sink if at least $r_{i}$  arcs are directed to itself from each of its neighbours. In the SMER dynamics, each $r$-sink node $i$ operates by reversing $r_{i}$ arcs towards all of its neighbours, next a new set of $r$-sinks operates in turn, and so on. Similarly to sinks under SER, only $r$-sink nodes are allowed to operate under SMER. Unlike SER, nodes may operate more than once consecutively in SMER dynamics.

Let $\mu_0, \mu_1,\ldots$ be the sequence of orientations produced by SMER over $\cM$ from the initial orientation $\mu_0$. As infinite sequences are of our interest (originally motivated by the Dining Philosophers with rates (DPPr) 
problem \cite{FRA}), let $a_{s}^{ij}$ denote the greatest multiple of $\gcd(r_i,r_j)$ of $r_i$ and $r_j$, which does not exceed the number of edges oriented from $i$ to $j$ in $\mu_s, s \ge 0$. Orientations $\mu_s$, such that 
$f_{ij} = a_{s}^{ij} + a_{s}^{ji}$, $s \geq 0$, remaining constant as a consequence of the two terms changing by a certain multiple of gcd$(r_i,r_j)$ (arcs reversed between neighbouring nodes $i$ and $j$). Let $\cM^{i,j}$ be the submultigraph 
of $\cM$ induced by a pair of neighbouring nodes $i$ and $j$, and let $\mu^{ij}_{0}$, $\mu^{ij}_{1},\ldots$\,
be the sequence of orientations of $\cM^{i,j}$ produced by SMER from $\mu^{ij}_{0}$. The following Lemma~\ref{lem:relations} states a basic topology constraint towards the definition of the multigraph $\cM$.

\begin{figure}[!ht]
\centering
\includegraphics[scale=0.5]{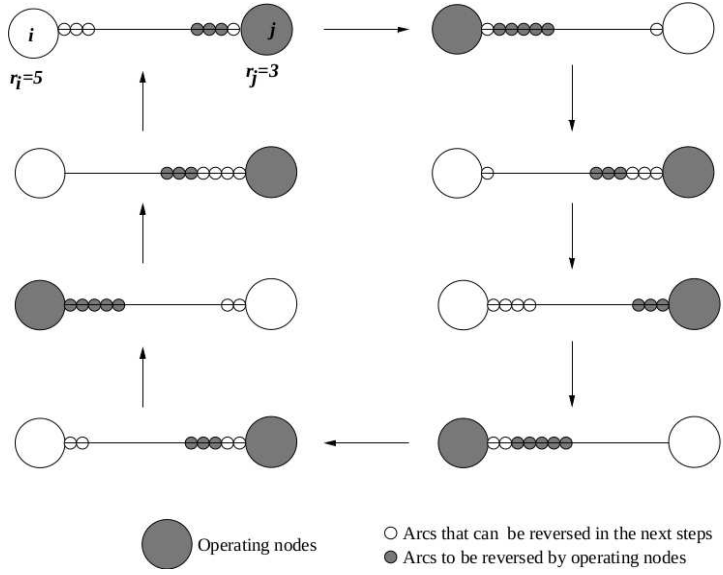}
\caption{\em An example of SMER, with period $P_{i,j} = 8$. Oriented arcs are represented by tokens.}
\label{fig:smer}
\end{figure}

\begin{lem} (\cite{FRA,fr:weig}) \label{lem:relations}
If $\max \{r_{i},r_{j}\} \leq e_{i,j} \leq r_{i} + r_{j} - 1$, aplication of SMER from $\mu^{ij}_{0}$ on $\cM^{i,j}$ solves the instance of DPPr given by neighbouring nodes $i$ and $j$, $r_{i}$ and $r_{j}$, if and only if 
$f_{ij} = r_{i} + r_{j} - \gcd(r_i,r_j)$. In this case, the sequence $\mu_{0}^{ij}, \mu_{1}^{ij},\ldots \mu_{s}^{ij}$ 
($s\ge 0$) includes all orientations of $\cM^{ij}$ that are legal for $i$ and $j$ given $\mu_{0}^{ij}$ in a given arbitrary multigraph $\cM$. If no deadlock arises for any initial orientation of the arcs between i and j, then
\[
\max \{r_{i},r_{j}\} \le e_{i,j} \le r_{i} + r_{j} - 1\ \quad \text{and}\ \quad f_{i,j} = r_{i} + r_{j} - gcd(r_{i},r_{j}).\]
\end{lem}
It is important to remark that there is always at least one SMER solution for any target system's topology having 
arbitrary pre-specified reversibilities at any of its nodes~\cite{fr:weig}. According to Lemma~\ref{lem:relations}, since 
$e_{i,j} = r_{i} + r_{j} - 1$, either $i$ or $j$ is in a $r$-sink condition, independently of $\mu_s, s \geq 0$.
It may also be seen that, between all pairs of neighbouring nodes $i$ and $j$ in $\cM$, any SMER dynamics produces \emph{one unique} period, given by the relation $P_{i,j} = (r_{i} + r_{j})/\gcd(r_{i},r_{j})$ ~\cite{FRA,FR94}. This periodic property of SMER can be observed in Fig.~\ref{fig:smer}, where $P_{i,j} = 8$ and the nodes in $\cM$ share values that are pairwise coprime integers: such pairs $(r_i,r_j)$ have no common divisors (except $1$).

\section{The Distributed Wheel Sieve Algorithm using SMER}
Let $\cM = (V,\cE)$ be an arbitrary multigraph having $N$ nodes. For the sake of simplicity, the distributed algorithm is actually assumed to sieve the restricted interval $\{2\}\bigcup \{$odd integers in $[3,\ldots, n]$\}, according to the parity of $n$. Such a SMER-based sieving algorithm is called {\em Semi-SMER}; this in contrast with the SMER dynamics described in Section~3, which considers the whole neighbourhood of any given node.

The procedure {\em Semi-SMER} is designed for any current node process $i\in V$, and it uses local variables, defined as follows:
\begin{itemize}
\item The interval $I$ is set to an exclusive value within $\{1,7,11, 13,17,19,23,29\}$. For example, when we make use of the third wheel $\cW_{3}$ in the algorithm, the interval $J$ is set to a value of $\Pi_{3}$ extended to 
$30 = 2\times 3\times 5$, $60 = 2\times 3^2\times 5$ and $90 = 2\times 3^3\times 5$. 

\item \textit{Neigh}$_i$ denotes the set of neighbours of process $i$, and the number of incoming arcs 
oriented from every $j\in$ \textit{Neigh}$_i$ to the current process $i$ is denoted by the variable \textit{incoming}$_i[j]$;

\item $r_{i}[j]$ denotes the required number of arcs that shall be reversed by $i$ towards every 
$j\in$ \textit{Neigh}$_i$, independently. The variable $r_{i}[j]$ takes its values in the interval $I$, and the variable 
$r_{j}[i]$ takes its values in the interval $J$;

\item $e_{i}[j]$ denotes the number of undirected edges (both outgoing and incoming arcs) connecting every pair of neighbours $(i,j)$ in $\cM$ (see Fig.~\ref{fig:smer});
    
\item $a_{i}[j]$ denotes the number of incoming arcs oriented from each $j\in$ \textit{Neigh}$_i$ to $i$ 
in the initial orientation;
    
\item Process $i$ also maintains the boolean variables $rev\_arc_{i}[j]$ and $end\_period\_{i}[j]$. If, at the end of the {\em Semi-SMER} period, $rev\_arc_{i}[j]$ is true for $j\in$ \textit{Neigh}$_i$, then $r_{i}[j]$ and $r_{j}[i]$ are coprime numbers ($(r_{i},r{j}) = 1$). The value of $end\_period\_{i}[j]$ checks whether the {\em Semi-SMER} between two nodes ended its execution or not;

\item $PseudoPrimes$ contains the numbers generated by the extended wheel that consists in the remaining prime numbers.
\end{itemize} 
   
\newpage
\bigskip \bigskip \no \rule{8.3cm}{0.4pt}\\
\no {\bf Procedure\ {\em WheelSieve-SMER$(N)$}}\\[.3\baselineskip]
\no {\textbf{var}}\\
\indent $P_{i,j} = 0$ \hfill {\small ($\star$ $P_{i,j}$ contains the size of the period of the SMER between two nodes $\star$)}\\
\indent $\cP$; \hfill {\small ($\star$ $\cP$ is the set of the first $k$ prime numbers $\star$)}\\ 
\indent $PseudoPrimes = 0$;\\
\indent $p_{k+1}$; \hfill {\small ($\star$ $p_{k+1}$ is initialized with the next prime number $\star$)}\\ 
\indent $prime$: boolean\ {\textbf{init}} true;\\
\indent $incoming_{i}[j]$: integer;\\
\indent $rev\_arc_{i}[j]$: boolean\ {\textbf{init}} false;\\
\indent $end\_period_{i}[j]$: boolean\ {\textbf{init}} false; \hfill 
{\small ($\star$ $r_{i}[j]$, $r_{j}[i]$ are initialized with the values of $I$, $J$, resp. $\star$)}\\[.4\baselineskip]
\no {\textbf{Begin}} \\
\indent \indent {\textbf{If}}\ $r_{i}[j] \le r_{j}[i]$\ \textbf{Then}\\
\indent \indent\indent $a_{i}[j] = r_{i}[j]$;\\
\indent \indent\indent $incoming_{i}[j] = r_{i}[j]$;\\
\indent \indent\indent $e_{i}[j] = r_{i}[j] + r_{j}[i]-1$;\\
\indent \indent{\textbf{Else}}\\
\indent \indent\indent $a_{i}[j] = r_{i}[j] - 1 $;\\
\indent \indent\indent $incoming_{i}[j] = a_{i}[j]$;\\
\indent \indent\indent $e_{i}[j] = r_{i}[j]+r_{j}[i]-1$;\\
\indent \indent{\textbf{EndIf}}\\[.3\baselineskip] 
\indent \indent{\textbf{While not}}\ $end\_period_{i}[j]$ \\
\indent \indent\indent {\textbf{If}}\ $incoming_{i}[j] \ge r_{i}[j]$\ \textbf{Then}\ %
send message $\langle r_{i}[j]\rangle$ to $j \in Neigh_{i}$;\\
\indent \indent\indent \indent $incoming_{i}[j] = incoming_{i}[j] - r_{i}[j] $;\\
\indent \indent\indent \indent $P_{i,j} = P_{i,j} +1$; \hfill {\small ($\star$ The flipping arcs process is triggered $\star$)}\\ 
\indent \indent\indent {\textbf{Else}}\\
\indent \indent\indent \indent receive $\langle r_{j}[i]\rangle$ from $j \in Neigh_{i}$;\\
\indent \indent\indent \indent $r_{i}[j]= incoming_{i}[j] +  r_{j}[i] $;\\
\indent \indent\indent \indent $P_{i,j} = P_{i,j} +1$;\\
\indent \indent\indent {\textbf{EndIf}}\\
\indent \indent\indent {\textbf{If}}\ $incoming_{i}[j] = 0$ \textbf{Then}\ $rev\_arc_{i}[j] = true$;\ {\textbf{EndIf}} \\
\indent \indent\indent {\textbf{If}}\ $incoming_{i}[j] = a_{i}[j]$ \textbf{Then}\ $end\_period_{i}[j] = true$;\ 
{\textbf{EndIf}}\\
\indent \indent $PseudoPrimes = P_{i,j}$;\\
\indent \indent $end\_period_{i}[j]$: boolean {\textbf{init}} false; \hfill {\small ($\star$ $r_{i}[j]$ and $r_{j}[i]$ are initialized with the values}\\
\hspace*{9cm} {\small of $p_{k+1}$ and $PseudoPrimes$, resp. $\star$)}\\ 
\indent \indent {\textbf{If}}\ $r_{i}[j] \le r_{j}[i]$\ \textbf{Then}\\
\indent \indent\indent $a_{i}[j] = r_{i}[j]$;\\
\indent \indent\indent $incoming_{i}[j] = r_{i}[j]$;\\
\indent \indent\indent $e_{i}[j] = r_{i}[j] + r_{j}[i]-1$;\\
\indent \indent{\textbf{Else}}\\
\indent \indent\indent $a_{i}[j] = r_{i}[j] - 1 $;\\
\indent \indent\indent $incoming_{i}[j] = a_{i}[j]$;\\
\indent \indent\indent $e_{i}[j]= r_{i}[j]+r_{j}[i]-1$;\\
\indent \indent{\textbf{EndIf}}\\[.3\baselineskip] 
\indent \indent{\textbf{While not}}\ $end\_period_{i}[j]$\\
\indent \indent\indent {\textbf{If}}\ $incoming_{i}[j] \ge r_{i}[j]$\ \textbf{Then}\ send message $\langle r_{i}[j]\rangle$ to $j \in Neigh_{i}$;\\ 
\indent \indent\indent \indent $incoming_{i}[j] = incoming_{i}[j] - r_{i}[j]$; \hfill {\small ($\star$ The flipping arcs process is triggered $\star$)}

\s \no \indent \indent\indent {\textbf{Else}}\ receive $\langle r_{j}[i]\rangle$ from $j \in Neigh_{i}$;\\
\indent \indent\indent \indent $r_{i}[j]= incoming_{i}[j] +  r_{j}[i] $;\\
\indent \indent\indent {\textbf{EndIf}}\\
\indent \indent\indent {\textbf{If}}\ $incoming_{i}[j] = 0$ \textbf{Then}\ $rev\_arc_{i}[j] = true$;\ {\textbf{EndIf}}\\
\indent \indent\indent {\textbf{If}}\ $incoming_{i}[j] = a_{i}[j]$ \textbf{Then}\ $end\_period_{i}[j] = true$;\ 
{\textbf{EndIf}}\\
\indent \indent{\textbf{EndWhile}}\\
\indent {\textbf{If}}\ $rev\_arc_{i}[j] = true$\ \textbf{Then}\ $\cP \cup P_{i,j}$\ {\textbf{EndIf}}\\
\indent {\textbf{Return}}\ $\cP$ \hfill {\small ($\star$ $\cP$ is the set of primes in $[2\ldots, n]$ $\star$)}\\
\no {\textbf{End.}}\\
\rule{8.3cm}{0.4pt}

\m As pointed out, if we start initially the sieve with the third wheel there are eight processes, whose 
values are the numbers $\{1, 7, 11, 13, 17, 19, 23, 29\}$, that represent the values of $I$. The set $J$ 
is started in accordance with $N$; for example, if $N=230$, eight processes are needed, one for each value 
$\in J$. These values represent the multiples of $\Pi_{3}=30$. Beginning with these values (we consider that at
the beginning, each process knows its identity), and after executing the {\em WheelSieve-SMER}, we obtain a set ($PseudoPrimes$) composed with the values of all periods spread in between the values of $I$ and $J$. 
There remains the operation of sieving the $PseudoPrimes$ set with $p_{k+1}$ in order to obtain $\cP$.

\section{Worst-Case complexity of the Algorithm}
In order to sieve all primes from the interval $[2\ldots, n]$, the only fundamental operations explicitly used 
in the algorithm {\em Semi-SMER} are comparisons, additions and the sending and receiving of messages (arc reversals). 
Besides, a send-receive event and one comparison operation are assumed to take $\cO(1)$ number of time slots.

\m The number of steps required by {\em Semi-SMER} is proportional to the period involved between any two nodes 
of $\cM$ during the algorithm. Now, the largest period $P_{i,j}$ follows from Lemma~\ref{lem:relations} and~\cite{FRA}: 
$P_{i,j} = r_{i} + r_{j}$, when $(r_{i},r_{j}) = 1$. Since $r_{i}\le \lfloor \sqrt{n}\rfloor$ and $r_{j}\le n$, for any pair of nodes $(i,j)$, the procedure {\em Semi-SMER($n$)} requires at most $n + \lfloor \sqrt{n}\rfloor$ steps.

Similarly, for any current pair of nodes $(i,j)$ of $\cM$ smaller than $\sqrt{n}$, the number of messages exchanged in the \textbf{while} loop is proportional to $P_{i,j}\times \mathrm{deg}_{i}$, with $\mathrm{deg}_{i} = \#$\textit{Neigh}$_i$. Hence, if we let $P := \sup_{(i,j)\in V^2} P_{i,j}$ denote the largest period between all pairs $(i,j)\in V^2$, the maximum message complexity of the algorithm is proportional to $P\times \Delta_N$, where $1\le \Delta_N \le N - 1$ is the maximum multidegree of $\cM$. Finally, the message complexity achieves at most $n\Delta_N + \lfloor \sqrt{n}\rfloor \Delta_N$. The maximum amount of memory space required per process is $\cO(n)$ bits. 

\section{Conclusion and Perspectives}
This paper introduced a totally new kind of SMER-based distributed sieve algorithm that generates all primes 
in a given interval $[2\ldots, n]$. Apart from observing that the fundamental operation of the {\em Semi-SMER} algorithm is a local comparison, it is also worth noticing that no $gcd$ computation is needed. Moreover, no precomputation is assumed in the {\em Semi-SMER} complexity analysis (precomputation would take $\cO\b(n\log\log \Pi_{k}\b)$, where $\Pi_{k}$ denotes the product of the first $k$ prime numbers, in the wheel sieve). This approach seems also general enough to compute some of the elementary arithmetic functions in number theory. For instance, using the gcd and inverse, the least common multiple of integers, and various basic multiplicative arithmetic functions, e.g. Euler's totient function $\phi(n)$, M\"obius function $\mu(n)$ and divisor functions: $d(n)$, $\sigma(n)$, $\omega(n)$, $\Omega(n)$, etc.

\m Finally, it stems also from both computer-driven and theoretical results that the number of steps $T(n)$ executed 
by the algorithm stays always ``very close'' to the maximal number of steps. 
More precisely, $T(n) = n + \lfloor \sqrt{n}\rfloor - \varphi(n)$, where $\varphi(n)$ is a positive non periodic arithmetic function with rather small fluctuations when $n\ge 4$: we conjecture that $\varphi(n) < 5$ for ``almost every'' $n\ge 4$. Hence, for every $n\ge 4$, $\varphi(n)$ should yield an expected $\overline{\varphi(n)} = 2.47\ldots \pm\varepsilon_n$ for all $0\le \varepsilon_n < 1$, and the \emph{average} number of steps required by the algorithm should then be expected to achieve  $\overline{T(n)}\approx n + \lfloor \sqrt{n}\rfloor - 2.47\ldots$.

\bibliographystyle{article}
\def\bibfmta#1#2#3#4{ {\sc #1.} {#2}, \emph{#3} #4.}
\bibliographystyle{book}
\def\bibfmtb#1#2#3#4{ {\sc #1.} \emph{#2}, {#3}, #4.}

\end{document}